\documentclass[10pt,prc,nofootinbib,amsmath,amssymb,superscriptaddress,groupedaddress,aps]{revtex4-2}
\usepackage{dcolumn}
\usepackage{bm}
\usepackage{amssymb}
\usepackage{comment}
\usepackage{amsmath}
\usepackage{array}
\usepackage{subcaption}
\usepackage{graphicx}
\usepackage{nameref}

\usepackage[english]{babel}
\usepackage[autostyle, english=american]{csquotes}
\MakeOuterQuote{"}
\usepackage[dvipsnames,table]{xcolor}

\usepackage{etoolbox}

\usepackage[%
  colorlinks = true,
  citecolor  = RoyalBlue,
  linkcolor  = RoyalBlue,
  urlcolor   = RoyalBlue,
  unicode,
  ]{hyperref}

\begin{document}
\title{Jet Radius Dependence of Energy Loss in Pb+Pb Collisions: A Comparative Analysis of the Ratio of Nuclear Modification Factors and Fractional Energy Loss}

\author{Rafet Kavak}
\email{rafet.kavak@cern.ch}
\affiliation{\small{\it Department of Physics, University of Illinois, Urbana, IL 61801, USA}}

\author{Riccardo Longo}
\email{riccardo.longo@cern.ch}
\affiliation{\small{\it Department of Physics, University of Illinois, Urbana, IL 61801, USA}}
\affiliation{\small{\it Dipartimento di Fisica, Università and INFN Torino, Torino, TO 10125, Italia}}

\author{Anne M. Sickles}
\email{sickles@illinois.edu}
\affiliation{\small{\it Department of Physics, University of Illinois, Urbana, IL 61801, USA}}

\date{\today}
\begin{abstract}

The quark-gluon plasma (QGP) is a deconfined state of strongly interacting matter formed at extreme temperature and energy density in ultra-relativistic nucleus-nucleus collisions at RHIC and the LHC. High transverse momentum jets, produced in initial hard scatterings, traverse the QGP and lose energy via elastic and radiative processes—an effect known as jet quenching. The nuclear modification factor, $R_{\mathrm{AA}}$, defined as the ratio of the Pb+Pb jet yield to the $pp$ cross section scaled by the nuclear thickness function, is widely used to quantify jet quenching. However, its value depends strongly on both the $pp$ jet spectral shape and the strength of the quenching, complicating comparisons across jet selections. The fractional energy loss, $S_{\text{loss}}$, quantifying the average medium-induced momentum shift of jets, is designed to mitigate this dependence. In central  Pb+Pb collisions at $\sqrt{s_{\mathrm{NN}}}=5.02~\mathrm{TeV}$, we compile and compare published ATLAS and ALICE measurements of jet suppression for inclusive single-jet and dijet selections across multiple jet radii, considering (i) the ratio of the nuclear modification factor at a given radius to that at a reference radius of 0.2, and (ii) the fractional energy loss. The radius dependence of this ratio differs between single-jet and dijet measurements, and between ATLAS calorimeter jets and ALICE charged-particle jets, reflecting differences in kinematic event selections and jet constituents. Expressing the results in terms of $S_{\text{loss}}$ allows direct, radius-differential comparisons across experiments with reduced sensitivity to the $pp$ spectral slope. Combining these approaches enables constraints on the radius dependence of jet modification that account for selection biases, and facilitates cross-experiment benchmarking of jet quenching models.

\end{abstract}
\maketitle

\section{Introduction}
\label{sec:intro}

Ultra-relativistic heavy-ion collisions at the Relativistic Heavy Ion Collider (RHIC) and the Large Hadron Collider (LHC) create extreme conditions of temperature and energy density, enabling the formation of a deconfined state of quarks and gluons known as the quark-gluon plasma (QGP)~\cite{Collins:1974ky, Shuryak:1980tp}. In this state, quarks and gluons, the fundamental constituents of hadrons, are no longer confined within color-neutral particles and instead behave collectively as a strongly coupled fluid~\cite{Romatschke:2017ejr}. Jets—collimated sprays of particles that originate from the fragmentation and hadronization of highly energetic quarks and gluons (collectively referred to as partons)—are produced very early in the collision via hard quantum chromodynamics (QCD) scatterings, typically before the formation of the QGP~\cite{Connors:2017ptx}. Because the production rates and kinematics of high transverse momentum\footnote{Transverse momentum is defined as $p_{\mathrm{T}} = p \sin\theta$, where $p$ is the particle momentum and $\theta$ is the polar angle with respect to the beam ($z$-)axis. The pseudorapidity is $\eta = -\ln\tan(\theta/2)$, and the angular distance between two objects is $\Delta R = \sqrt{(\Delta\eta)^2 + (\Delta\phi)^2}$, where $\phi$ is the azimuthal angle around the $z$-axis.} ($p_{\mathrm{T}}$) partons can be calculated reliably within perturbative QCD~\cite{Marzani:2019hun}, jets in proton-proton ($pp$) collisions provide a baseline. Relative to this baseline, deviations of jet observables in nucleus-nucleus (A+A) collisions quantify medium-induced modifications, collectively known as jet quenching~\cite{Bjorken:1982tu, Gyulassy:1990ye}; see Ref.~\cite{Cunqueiro:2021wls} for a recent review.

A central and widely utilized observable is the nuclear modification factor, $R_{\mathrm{AA}}$, which quantifies the suppression of jet yields in A+A collisions relative to $pp$ collisions. It is defined as
\begin{equation}
    R_{\mathrm{AA}} = \frac{1}{N_{\mathrm{evt}}} \frac{\mathrm{d}^2 \mathrm{N}^{\mathrm{PbPb}}_\text{jets}}{\mathrm{d}p_{\mathrm{T}}\mathrm{d}\eta} \Bigg/ \left\langle T_{\mathrm{AA}} \right\rangle \frac{\mathrm{d}^2 \sigma^{pp}_\text{jets}}{\mathrm{d}p_{\mathrm{T}}\mathrm{d}\eta},
    \label{eq:raa_definition}
\end{equation}
where $N_{\mathrm{evt}}$ is the number of Pb+Pb events in a given centrality class, $\mathrm{d}^2 \mathrm{N}^{\mathrm{PbPb}}_\text{jets}/\mathrm{d}p_{\mathrm{T}}\mathrm{d}\eta$ is the differential jet yield in Pb+Pb collisions, $\mathrm{d}^2 \sigma^{pp}/\mathrm{d}p_{\mathrm{T}}\mathrm{d}\eta$ is the inclusive jet cross section in $pp$ collisions, and $\left\langle T_{\mathrm{AA}} \right\rangle$ is the nuclear overlap function derived from Glauber modeling~\cite{Miller:2007ri}. In the absence of nuclear effects, the $R_{\mathrm{AA}}$ is expected to equal unity, whereas an $R_{\mathrm{AA}}<1$ usually is taken as due to energy loss from interactions with the QGP.

The dependence of the $R_{\mathrm{AA}}$ on the jet radius $R$ probes the redistribution of energy within and outside the jet cone and varies with kinematics and event selection~\cite{Mehtar-Tani:2021fud, Apolinario:2022vzg, Mehtar-Tani:2024jtd}. Experimentally, CMS reports no significant $R$ dependence for $p_{\mathrm{T}}>400~\mathrm{GeV}$ in central Pb+Pb collisions~\cite{CMS:2021vui}; ALICE, in a lower $p_{\mathrm{T}}$ range ($p_{\mathrm{T}}<120~\mathrm{GeV}$), reports stronger suppression at larger $R$, most prominently at $R=0.6$~\cite{ALICE:2023waz}; ATLAS reports a difference between $R=0.4$ and $R=0.2$ over $100$–$800~\mathrm{GeV}$~\cite{ATLAS:2014ipv}. These contrasts motivate radius-differential observables that reduce correlated systematics.

Comparing the suppression of inclusive jets and jets in dijet events provides complementary insights into jet quenching mechanisms. Dijet events isolate the $2\!\to\!2$ partonic scattering—two approximately back-to-back partons produced in the initial collision that fragment into two jets—and, by requiring a recoil jet above a set $p_{\mathrm{T}}$ threshold, place an effective lower bound on the quenching of the subleading jet admitted to the sample~\cite{ATLAS:2010isq, CMS:2011iwn}. Recent ATLAS results highlight an important difference in the $R$-dependence between inclusive and dijet samples: in central Pb+Pb collisions at $\sqrt{s_{\mathrm{NN}}}=5.02$ TeV, the dijet pair nuclear modification factors $R_{\mathrm{AA}}^{\text{pair}}(p_{\mathrm{T},1})$ and $R_{\mathrm{AA}}^{\text{pair}}(p_{\mathrm{T},2})$ for leading and subleading jets increase with the jet radius (i.e. the suppression decreases for larger $R$)~\cite{ATLAS:2024jtu}.

To characterize the dependence of jet suppression on the jet radius parameter $R$, it is useful to construct a double ratio of nuclear modification factors, defined as
\begin{equation}
    R_{\mathrm{AA}}^{R/0.2} = \frac{R_{\mathrm{AA}}^R}{R_{\mathrm{AA}}^{0.2}},
    \label{eq:raa_double_ratio_simple}
\end{equation}
where $R_{\mathrm{AA}}^R$ denotes the nuclear modification factor for jets reconstructed with a given radius $R$, and $R=0.2$ is used as a reference. A value of $R_{\mathrm{AA}}^{R/0.2} > 1$ implies reduced suppression for larger-$R$ jets, while $R_{\mathrm{AA}}^{R/0.2} < 1$ indicates an increasing suppression with larger radii. 

While the $R_{\mathrm{AA}}$ and its double ratio provide valuable insight into suppression patterns and their dependence on jet parameters, they remain indirect observables, sensitive to both energy loss and the steeply falling $p_{\mathrm{T}}$ spectrum. A complementary approach is to directly estimate the shift in $p_{\mathrm{T}}$ from $pp$ collisions caused by jet quenching. This motivates the use of the fractional energy loss variable, $S_{\text{loss}}$~\cite{PHENIX:2004vcz, PHENIX:2006wwy}. The $S_{\text{loss}}$ variable quantifies the effective momentum loss of jets by comparing the Pb+Pb yield to the $pp$ spectrum at the same production rate. The method identifies, for each jet momentum $p_{\mathrm{T}}^{pp}$ in $pp$ collisions, the corresponding value in Pb+Pb collisions that yields the same rate after scaling by the $\langle T_{\mathrm{AA}} \rangle$. The formal derivation and the numerical implementation of this procedure are provided in Section~\ref{sec:methods}.

Unlike the $R_{\mathrm{AA}}$, which reflects the relative suppression of yields, $S_{\text{loss}}$ provides a direct estimate of the average momentum lost by the jets in the medium as a function of their initial energy. This observable is less sensitive to the shape of the $pp$ spectrum and offers improved interpretability when comparing different jet selections, such as inclusive versus photon-tagged jets, or reconstructed with different radii~\cite{ATLAS:2023iad}.

In this study, we perform a comparative analysis of jet quenching observables in central (0--10\%) Pb+Pb collisions at $\sqrt{s_{\mathrm{NN}}} = 5.02~\mathrm{TeV}$, utilizing published measurements from the ATLAS~\cite{ATLAS:2018gwx, ATLAS:2023hso, ATLAS:2024jtu} and ALICE~\cite{ALICE:2023waz} collaborations. All considered measurements reconstruct jets using the anti-$k_T$ algorithm~\cite{Cacciari:2008gp}. We systematically construct the double nuclear modification factor, $R_{\mathrm{AA}}^{R/0.2}$, and extract the fractional energy loss, $S_{\text{loss}}$, for a range of jet radii and selections—including inclusive, photon-tagged, charged particle jets, and dijets. By analyzing both ratio-based ($R_{\mathrm{AA}}^{R/0.2}$) and shift-based ($S_{\text{loss}}$) observables, we aim to disentangle the competing effects of energy redistribution and the steeply falling jet spectrum. This joint approach provides radius-differential, selection-specific constraints on jet-energy loss mechanisms in the QGP and enables robust, data-driven benchmarking of jet quenching models across experiments.

\section{Methods}
\label{sec:methods}

\subsection{Double Nuclear Modification Factor \texorpdfstring{$R_{\mathrm{AA}}^{R/0.2}$}{RAA} Construction}

In this analysis, $R_{\mathrm{AA}}^{R/0.2}$ values are extracted directly from ALICE published charged-particle jet~\cite{ALICE:2023waz} and ATLAS dijet results~\cite{ATLAS:2024jtu}, or constructed from published single-radius $R_{\mathrm{AA}}$ values when double ratios are not provided. For ATLAS inclusive jets, the double ratio is determined by taking the ratio of the reported $R_{\mathrm{AA}}$ for $R=0.4$ jets from Ref.~\cite{ATLAS:2018gwx} and for $R=0.2$ jets from Ref.~\cite{ATLAS:2023hso}, both evaluated in the 0--10\% centrality class at $\sqrt{s_{\mathrm{NN}}} = 5.02~\mathrm{TeV}$. Statistical and systematic uncertainties are propagated independently in the numerator and denominator, assuming no correlations, and combined in quadrature to obtain the total uncertainty on the double ratio.

\subsection{Jet Yield Ratios at Different Radii} \label{sec:yield_ratio_def}

It is useful to express $R_{\mathrm{AA}}^{R/0.2}$ in terms of ratios of jet yields at fixed $p_{\mathrm{T}}$, since this makes explicit that the double ratio compares how the jet yield changes with radius in Pb+Pb collisions relative to the corresponding change in $pp$ collisions and connects directly to the jet cross-section ratios measured by ATLAS and ALICE.\footnote{Using the definition of $R_{\mathrm{AA}}$ in Eq.~\eqref{eq:raa_definition}, one has $R_{\mathrm{AA}}^{R}(p_{\mathrm{T}}) = \big[(1/N_{\mathrm{evt}})\,\mathrm{d}\mathrm{N}^{\mathrm{PbPb}}_{R}/\mathrm{d}p_{\mathrm{T}}\big] \big/ \big[\langle T_{\mathrm{AA}}\rangle\,\mathrm{d}\sigma^{pp}_{R}/\mathrm{d}p_{\mathrm{T}}\big]$. In the double ratio $R_{\mathrm{AA}}^{R}/R_{\mathrm{AA}}^{0.2}$, the factors $1/N_{\mathrm{evt}}$ and $\langle T_{\mathrm{AA}}\rangle$ cancel between numerator and denominator, leaving a ratio of Pb+Pb yields divided by a ratio of $pp$ cross sections at fixed $p_{\mathrm{T}}$. Writing the $pp$ cross sections as per-event yields, $\mathrm{d}\sigma^{pp}_{R}/\mathrm{d}p_{\mathrm{T}} = (1/L_{pp})\,\mathrm{d}\mathrm{N}^{pp}_{R}/\mathrm{d}p_{\mathrm{T}}$, the integrated luminosity $L_{pp}$ cancels as well. Thus the yield ratios in Eq.~\eqref{eq:raa_double_ratio_yields} are numerically equivalent to the cross-section ratios used in the ALICE~\cite{ALICE:2023waz} and ATLAS~\cite{ATLAS:2024jtu} analyses.} 

Using per-event jet yields for a given jet radius $R$, the double ratio in Eq.~\eqref{eq:raa_double_ratio_simple} can be written as
\begin{equation}
  R_{\mathrm{AA}}^{R/0.2}(p_{\mathrm{T}}) = \frac{R_{\mathrm{AA}}^{R}(p_{\mathrm{T}})}{R_{\mathrm{AA}}^{0.2}(p_{\mathrm{T}})} = \frac{\dfrac{\mathrm{d}\mathrm{N}^{\mathrm{PbPb}}}{\mathrm{d}p_{\mathrm{T},\mathrm{jet}}}(R) \Bigg/ \dfrac{\mathrm{d}\mathrm{N}^{\mathrm{PbPb}}}{\mathrm{d}p_{\mathrm{T},\mathrm{jet}}}(0.2)}{\dfrac{\mathrm{d}\mathrm{N}^{pp}}{\mathrm{d}p_{\mathrm{T},\mathrm{jet}}}(R) \Bigg/ \dfrac{\mathrm{d}\mathrm{N}^{pp}}{\mathrm{d}p_{\mathrm{T},\mathrm{jet}}}(0.2)}.
  \label{eq:raa_double_ratio_yields}
\end{equation}
In this form, the building blocks are ratios of jet yields at two radii evaluated at the same $p_{\mathrm{T}}$, which we refer to as \emph{yield ratios}. As for the $R_{\mathrm{AA}}^{R/0.2}$ values, the yield ratios are taken directly from the published jet cross-section ratios reported by ALICE~\cite{ALICE:2023waz} and ATLAS~\cite{ATLAS:2024jtu} and are interpreted as yield ratios according to Eq.~\eqref{eq:raa_double_ratio_yields}. In the results section, we present the Pb+Pb and $pp$ yield ratios separately, which correspond to the numerator and denominator of Eq.~\eqref{eq:raa_double_ratio_yields}, respectively.

\subsection{Extraction of Fractional Energy Loss \texorpdfstring{$S_{\text{loss}}$}{Sloss}}

$S_{\text{loss}}$ is extracted by comparing jet spectra measured in A+A and $pp$ collisions, following the numerical approach originally proposed by the PHENIX Collaboration~\cite{PHENIX:2015vqa}. The central idea is to treat the $pp$ jet spectrum, scaled by the nuclear overlap function $\langle T_{\mathrm{AA}}\rangle$ for the corresponding Pb+Pb centrality class, as a reference baseline. To facilitate interpolation and yield-matching, the scaled $pp$ spectrum is parametrized using a modified power-law function~\cite{Spousta:2015fca} of the form:
\begin{equation}
    \frac{\mathrm{d}^2\sigma_{pp}}{\mathrm{d}p_{\mathrm{T}}d\eta} = c_1 \cdot \left( \frac{c_2}{p_{\mathrm{T}}} \right)^{c_3 + c_4 \cdot \log\left( \frac{p_{\mathrm{T}}}{c_2} \right)},
\label{eq:modified_powerlaw}
\end{equation}
where $c_i$ represent the fit parameters determined from data. For inclusive jets, an additional $c_5 \cdot p_{\mathrm{T}}$ term is included in the exponent of Eq.~\eqref{eq:modified_powerlaw} to better describe the shape of the steeply falling spectra at high $p_{\mathrm{T}}$.

For each $p_{\mathrm{T}}$ bin in the Pb+Pb measurement, the corresponding transverse momentum $p_{\mathrm{T}}^{pp}$ in the fitted and scaled $pp$ spectrum that yields the same jet production rate is determined numerically. This corresponds to finding the horizontal shift, $\Delta p_{\mathrm{T}}$, required to align the $pp$ reference spectrum with the Pb+Pb yield, defined as:
\begin{equation}
    \Delta p_{\mathrm{T}} = p_{\mathrm{T}}^{pp} - p_{\mathrm{T}}^{\mathrm{PbPb}},
\end{equation}
where $p_{\mathrm{T}}^{\mathrm{PbPb}}$ is the momentum at which the scaled Pb+Pb yield matches the interpolated $pp$ yield. This matching condition can be expressed as:
\begin{equation}
    \frac{1}{\langle T_{\mathrm{AA}} \rangle} \frac{1}{N_{\mathrm{evt}}}   \left. \frac{\mathrm{d}^2 \mathrm{N}^{\mathrm{PbPb}}}{\mathrm{d}p_{\mathrm{T}}^{\mathrm{PbPb}}\,\mathrm{d}\eta} \right|_{p_{\mathrm{T}}^{\mathrm{PbPb}} = p_{\mathrm{T}}^{pp} - \Delta p_{\mathrm{T}}} = \left. \frac{\mathrm{d}^2 \sigma^{pp}}{\mathrm{d}p_{\mathrm{T}}^{pp}\,\mathrm{d}\eta} \right|_{p_{\mathrm{T}}^{pp}}.
    \label{eq:sloss_matching}
\end{equation}

This work follows the original PHENIX prescription~\cite{PHENIX:2015vqa}: we solve Eq.~\eqref{eq:sloss_matching} bin by bin using the measured Pb+Pb spectrum and the $T_{\mathrm{AA}}$-scaled $pp$ reference, and we do not introduce an additional Jacobian factor.\footnote{In the ATLAS formulation~\cite{ATLAS:2023iad}, a Jacobian term $\left(1 + \frac{\mathrm{d}\Delta p_{\mathrm{T}}}{\mathrm{d}p_{\mathrm{T}}^{pp}}\right)$ is multiplied to the right-hand side of Eq.~\eqref{eq:sloss_matching} to preserve the total number of jets. Since ATLAS fits the Pb+Pb spectra, the derivative $\mathrm{d}\Delta p_{\mathrm{T}}/\mathrm{d}p_{\mathrm{T}}^{pp}$ can be evaluated consistently over the full $p_{\mathrm{T}}$ range. In the present analysis, we work directly with the measured Pb+Pb points and do not perform such a fit, so we omit this term.} Finally, $S_{\text{loss}}$ is computed as:
\begin{equation}
    S_{\text{loss}}(p_{\mathrm{T}}^{pp}) = \frac{\Delta p_{\mathrm{T}}}{p_{\mathrm{T}}^{pp}},
\label{eq:sloss_definition}
\end{equation}
which characterizes the relative magnitude of momentum loss experienced by jets traversing the medium.

The uncertainties on $\Delta p_{\mathrm{T}}$ and $S_{\text{loss}}$ are estimated by propagating the experimental uncertainty on the reported nuclear modification factor $R_{\mathrm{AA}}$, which already accounts for correlations between the $pp$ and Pb+Pb measurements. For each $p_{\mathrm{T}}$ bin, the central value of $R_{\mathrm{AA}}$ is used to reconstruct the nominal Pb+Pb yield, while the reported uncertainties are used to generate upper and lower yield variations according to
\begin{equation}
    R_{\mathrm{AA}}(p_{\mathrm{T}}) \cdot \frac{\mathrm{d}^2 \sigma_{pp}}{\mathrm{d}p_{\mathrm{T}}\,\mathrm{d}\eta} = \frac{1}{\langle T_{\mathrm{AA}} \rangle N_{\mathrm{evt}}} \frac{\mathrm{d}^2 \mathrm{N}_{\mathrm{PbPb}}}{\mathrm{d}p_{\mathrm{T}}\,\mathrm{d}\eta}.
\end{equation}
Three yield histograms—nominal, upper, and lower—are then fitted independently, and the $S_{\text{loss}}$ extraction procedure is repeated for each case. The resulting variation in $S_{\text{loss}}$ is taken as the systematic uncertainty associated with the experimental inputs.

\subsection{Summary of Data Sources and Calculations}

To ensure clarity regarding which observables are newly calculated in this work and which are directly adopted from published results, Table~\ref{tab:inputs_summary} summarizes the relevant datasets and methodologies. 

\begin{table}[hbt!]
  \centering
  \caption{Summary of observables, their sources, and methods of derivation.}
  \label{tab:inputs_summary}
  \begin{tabular}{lll}
      \hline
      \textbf{Observable} & \textbf{Source} & \textbf{Method} \\
      \hline
      $R_{\mathrm{AA}}^{R/0.2}$ (dijets) & ATLAS~\cite{ATLAS:2024jtu} & Replotted from published values \\
      $R_{\mathrm{AA}}^{R/0.2}$ (charged particle jets) & ALICE~\cite{ALICE:2023waz} & Replotted from published values \\
      $R_{\mathrm{AA}}^{R/0.2}$ (inclusive jets) & ATLAS~\cite{ATLAS:2018gwx, ATLAS:2023hso} & Constructed by dividing published $R_{\mathrm{AA}}$ values at different radii \\
      $\dfrac{\mathrm{d N}^R}{\mathrm{d} p_{\mathrm{T}, \mathrm{jet}}} \Big/ \dfrac{\mathrm{d N}^{0.2}}{\mathrm{d} p_{\mathrm{T}, \mathrm{jet}}}$ (dijets) & ATLAS~\cite{ATLAS:2024jtu} & Replotted from published values \\
      $\dfrac{\mathrm{d N}^R}{\mathrm{d} p_{\mathrm{T}, \mathrm{jet}}} \Big/ \dfrac{\mathrm{d N}^{0.2}}{\mathrm{d} p_{\mathrm{T}, \mathrm{jet}}}$ (charged particle jets) & ALICE~\cite{ALICE:2023waz} & Replotted from published values \\
      $S_{\text{loss}}$ (dijets) & This work & Calculated using fit-based matching procedure \\
      $S_{\text{loss}}$ (charged-particle jets) & This work & Calculated using fit-based matching procedure \\
      $S_{\text{loss}}$ ($R=0.2$ inclusive jets) & This work & Calculated using fit-based matching procedure \\
      $S_{\text{loss}}$ ($R=0.4$ inclusive jets) & ATLAS~\cite{ATLAS:2023iad} & Replotted from published values \\
      $S_{\text{loss}}$ ($\gamma$-tagged jets) & ATLAS~\cite{ATLAS:2023iad} & Replotted from published values \\
      \hline
  \end{tabular}
\end{table}

\section{Results}
\label{sec:results}
This section presents a comparative analysis of jet quenching observables across different jet reconstruction radii in central (0--10\%) Pb+Pb collisions at $\sqrt{s_{\mathrm{NN}}} = 5.02~\mathrm{TeV}$. We focus on three key quantities: the double nuclear modification factor $R_{\mathrm{AA}}^{R/0.2}$, yield ratios in $pp$ and Pb+Pb collisions, and the fractional energy loss $S_{\text{loss}}$.

To enable a more meaningful comparison between charged-particle jets measured by ALICE and fully reconstructed calorimeter jets from ATLAS, the ALICE results are scaled by a factor of 
$3/2$ in jet $p_{\mathrm{T}}$. This factor accounts for the typical charged fraction of inclusive jets and approximates the missing neutral energy component, consistent with the ALICE estimate that, for pion-dominated final states, the charged-jet momentum is approximately 2/3 of the full jet momentum~\cite{ALICE:2023ama}.

\subsection{Double Nuclear Modification Factor \texorpdfstring{$R_{\mathrm{AA}}^{R/0.2}$}{RAAr02}}

\begin{figure}[hbt!]
  \centering
  \includegraphics[width=0.98\textwidth]{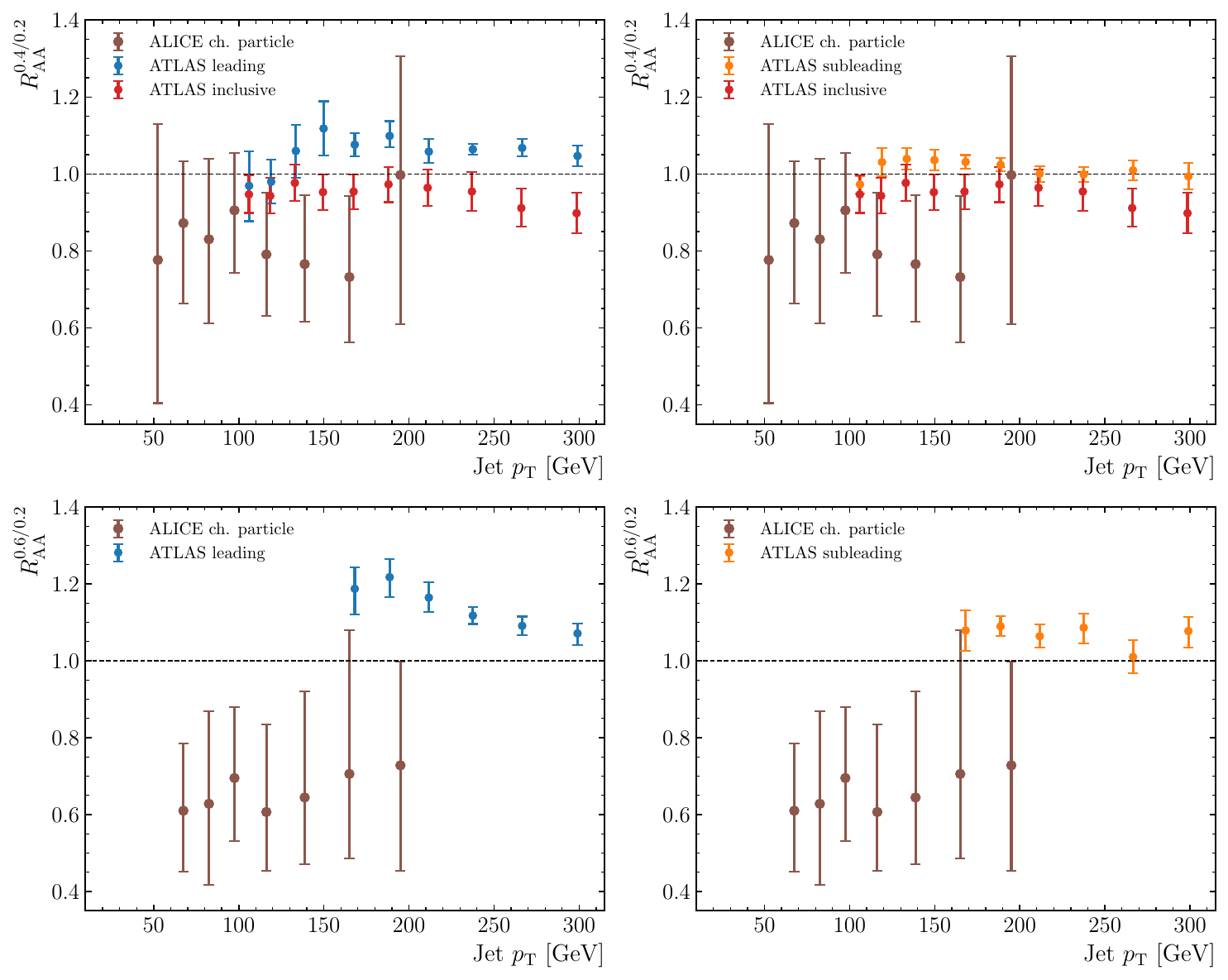}
  \caption{Double nuclear modification factor $R_{\mathrm{AA}}^{R/0.2}$ as a function of jet $p_{\mathrm{T}}$ in central (0--10\%) Pb+Pb collisions at $\sqrt{s_{\mathrm{NN}}} = 5.02~\mathrm{TeV}$. Results are shown for $R=0.4$ (top row) and $R=0.6$ (bottom row), using data from ALICE charged-particle jets~\cite{ALICE:2023waz}, ATLAS inclusive jets~\cite{ATLAS:2018gwx}, and ATLAS dijets (leading and subleading jets)~\cite{ATLAS:2024jtu}. 
  For consistency with full jet energy, the ALICE data are scaled by a factor of $3/2$ in jet $p_{\mathrm{T}}$. Error bars represent the combined statistical and systematic uncertainties, added in quadrature.}
  \label{fig:ratios_of_raa}
\end{figure}

To investigate how jet suppression varies with the reconstruction radius, we first examine $R_{\mathrm{AA}}^{R/0.2}$ as a function of jet $p_{\mathrm{T}}$.
Figure~\ref{fig:ratios_of_raa} presents $R_{\mathrm{AA}}^{R/0.2}$ for jets reconstructed with $R=0.4$ (top) and $R=0.6$ (bottom). The panels include results from ALICE charged-particle jets~\cite{ALICE:2023waz}, ATLAS inclusive jets~\cite{ATLAS:2018gwx}, and ATLAS dijets~\cite{ATLAS:2024jtu}, with the latter shown separately for leading and subleading jets.

For $R=0.4$ jets, all three measurements are available. The ATLAS leading jets exhibit a clear enhancement above unity, with $R_{\mathrm{AA}}^{0.4/0.2}$ ranging from approximately $1.10 \pm 0.05$ to $1.18 \pm 0.08$ in the range $120 < p_{\mathrm{T}} < 300~\mathrm{GeV}$, indicating reduced suppression at a larger radius. The subleading jets show a smaller enhancement, typically around $1.05 \pm 0.07$. The ATLAS inclusive jet results are mildly below unity across the full $p_{\mathrm{T}}$ range, with $R_{\mathrm{AA}}^{0.4/0.2}$ values spanning $0.90 \pm 0.05$ to $0.95 \pm 0.05$, suggesting a modest residual suppression even at larger $R$. The ALICE charged-particle jet ratios remain below unity throughout, with $R_{\mathrm{AA}}^{0.4/0.2}$ ranging from $0.75 \pm 0.35$ at low $p_{\mathrm{T}}$ to $1.0 \pm 0.30$ at higher $p_{\mathrm{T}}$, suggesting a weak dependence on jet radius, or reflecting differences in jet definition and charged-energy composition.

For $R=0.6$ jets, only ALICE charged-particle and ATLAS dijet measurements are available. The ATLAS leading jets exhibit a strong enhancement, with $R_{\mathrm{AA}}^{0.6/0.2}$ ranging between $1.10 \pm 0.05$ and $1.20 \pm 0.10$, indicating reduced suppression at larger radius. Subleading jets from dijet pairs show ratios closer to unity, approximately $1.05 \pm 0.03$, across the measured $p_{\mathrm{T}}$ range. In contrast, the ALICE charged-particle jet results show a pronounced $R$-dependence with $R_{\mathrm{AA}}^{0.6/0.2}$ between $0.60 \pm 0.25$ and $0.70 \pm 0.35$ at both low and intermediate $p_{\mathrm{T}}$. This trend suggests a possible increase in suppression for larger-radius charged jets, although the uncertainties are substantial.

\subsection{Jet Yield Ratios at Different Radii}
\begin{figure}[hbt!]
  \centering
  \includegraphics[width=0.98\textwidth]{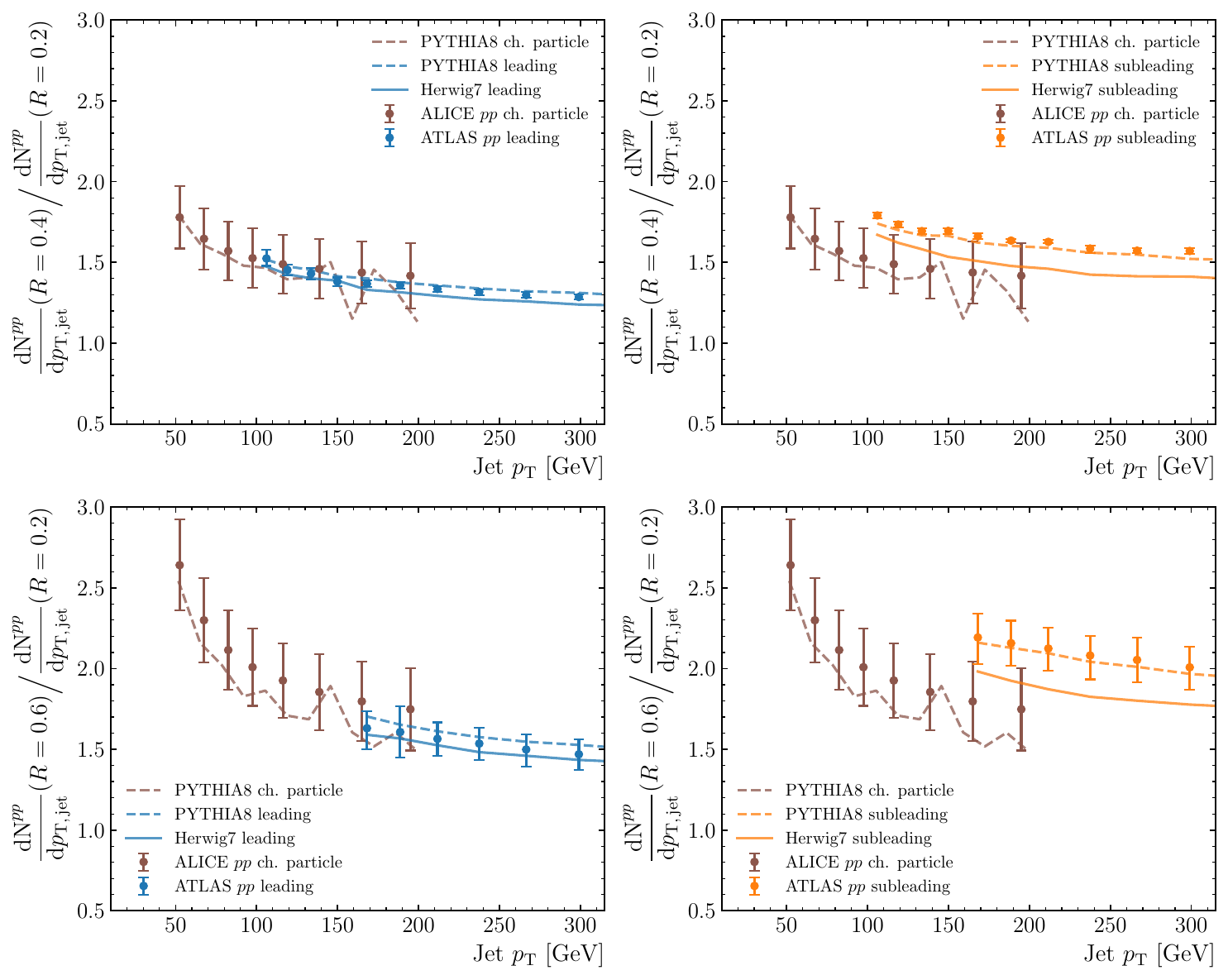}
  \caption{Ratio of jet yields in $pp$ collisions at $\sqrt{s} = 5.02~\mathrm{TeV}$, shown as a function of jet $p_{\mathrm{T}}$. The top row corresponds to $R = 0.4$, and the bottom row to $R = 0.6$. The left and right panels present ATLAS leading and subleading jets~\cite{ATLAS:2024jtu}, respectively. The ALICE charged-particle jet ratios~\cite{ALICE:2023waz} are overlaid in both panels for comparison. The ALICE jet $p_{\mathrm{T}}$ values are scaled by a factor of $3/2$ to approximate the total jet energy. Theoretical predictions from PYTHIA8~\cite{Sjostrand:2014zea} (dashed lines) and Herwig7~\cite{Bahr:2008pv, Bellm:2015jjp} (solid lines) are shown for comparison. Error bars represent the combined statistical and systematic uncertainties, added in quadrature.}
  \label{fig:ratios_of_cross_sec_pp}
\end{figure}

Figure~\ref{fig:ratios_of_cross_sec_pp} presents the jet yield ratios in $pp$ collisions at $\sqrt{s} = 5.02~\mathrm{TeV}$, which quantify the fractional increase in reconstructed jet yield when using larger radii $R = 0.4$ or $R = 0.6$ relative to the narrow $R = 0.2$ baseline. The figure includes experimental measurements from ALICE and ATLAS: ALICE results are shown for inclusive charged-particle jets, while ATLAS data are provided separately for leading and subleading jets in dijet events. Predictions from the PYTHIA8~\cite{Sjostrand:2014zea} and Herwig7~\cite{Bahr:2008pv, Bellm:2015jjp} Monte Carlo generators are also shown for comparison.

All datasets exhibit a decreasing trend of the yield ratio with increasing $p_{\mathrm{T}}$, indicating that jets become effectively narrower at higher momenta. For the ATLAS dijet sample, the subleading jets systematically have larger yield ratios than the leading jets over the measured $p_{\mathrm{T}}$ range, consistent with subleading jets having a broader angular structure. PYTHIA8 provides a good description of both the ALICE charged-particle and the ATLAS dijet results for leading and subleading jets, while Herwig7 predicts smaller ratios, most notably for the ATLAS subleading sample where it undershoots the data by typically 10-20\%. The comparison between ALICE charged-particle and ATLAS leading jets shows reasonable agreement for $R=0.4$ and $R=0.6$ jets within uncertainties. For subleading jets, the ALICE charged-particle yield ratios lie below the ATLAS values in the overlapping high-$p_{\mathrm{T}}$ region.

\begin{figure}[hbt!]
  \centering
  \includegraphics[width=0.98\textwidth]{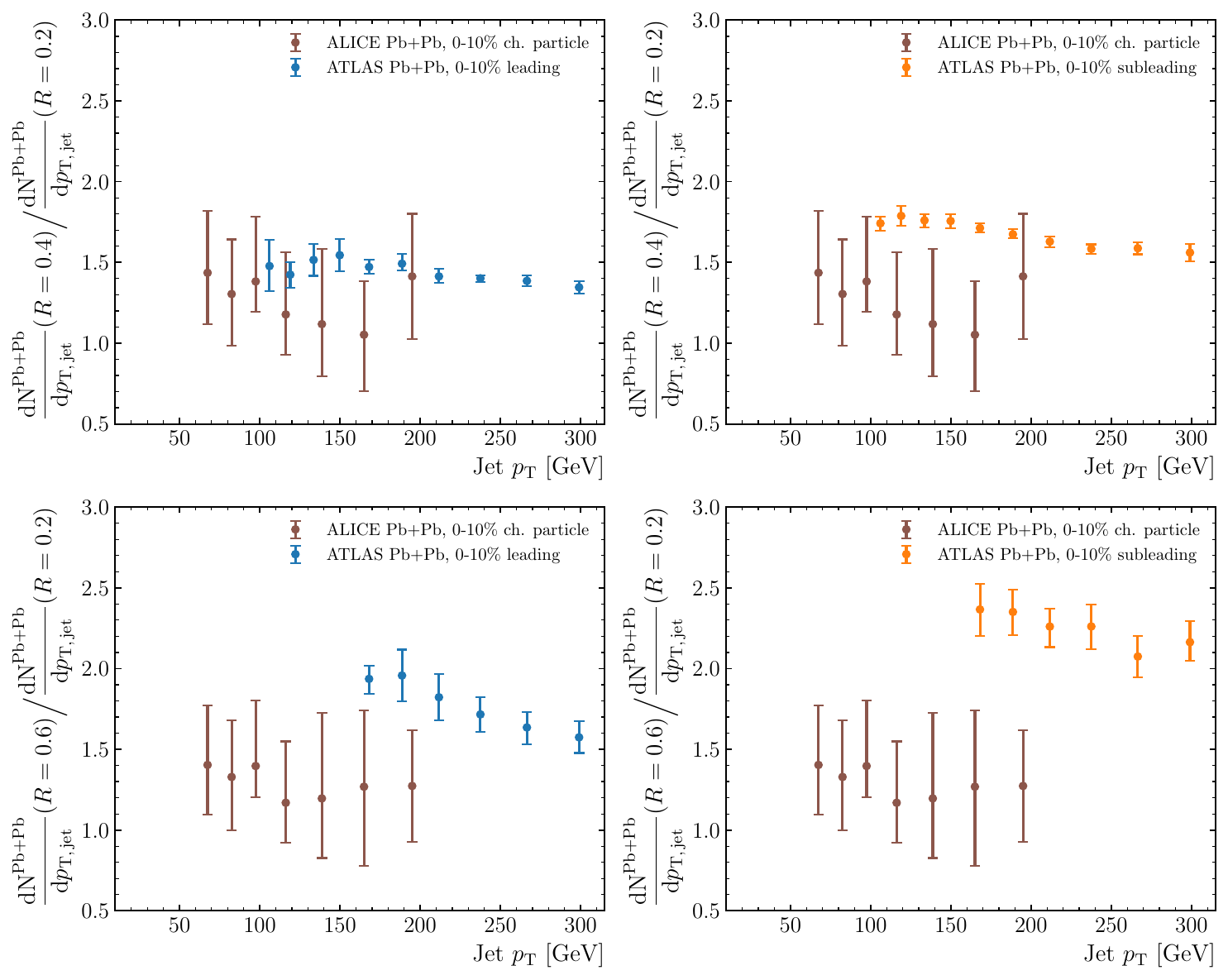}
  \caption{Ratio of jet yield in central (0--10\%) Pb+Pb collisions at $\sqrt{s_{\mathrm{NN}}} = 5.02~\mathrm{TeV}$, shown as a function of jet $p_{\mathrm{T}}$. The top row corresponds to $R = 0.4$ and the bottom row to $R = 0.6$; the left and right columns show leading and subleading jets, respectively. ALICE results are for charged-particle jets~\cite{ALICE:2023waz}, and ATLAS results are for fully reconstructed jets in dijet events~\cite{ATLAS:2024jtu}. ALICE jet $p_{\mathrm{T}}$ values are scaled by a factor of $3/2$ to approximate full jet energy. Error bars represent the sum in quadrature of statistical and systematic uncertainties.}
  \label{fig:ratios_of_cross_sec_pbpb}
\end{figure}

Figure~\ref{fig:ratios_of_cross_sec_pbpb} shows the corresponding jet yield ratios in central (0--10\%) Pb+Pb collisions. Compared to the $pp$ case, significant differences are observed in both the magnitude and trend of the ratios. 

In the $R = 0.4$ case, the ALICE results lie systematically below those of ATLAS, particularly in the subleading jet case, where discrepancies of up to 20\% are observed. This difference is larger than in the corresponding $pp$ results. At $R = 0.6$, the separation between the two experiments increases: for both leading and subleading jets the ALICE yield ratios lie below the corresponding ATLAS results over the common $p_{\mathrm{T}}$ interval, with deviations reaching about 20--25\%.

\subsection{Fractional Energy Loss Analysis}

To quantify the average momentum loss of jets due to interactions with the QGP, we extract the fractional energy loss, $S_{\text{loss}}$, from matched $pp$ and Pb+Pb jet spectra, as described in Section~\ref{sec:methods}. We begin by focusing exclusively on ATLAS results in the 0--10\% centrality class to isolate and compare the behavior of different jet types.

\begin{figure}[hbt!]
  \centering
  \begin{subfigure}{0.49\textwidth}
    \centering
    \includegraphics[width=\textwidth]{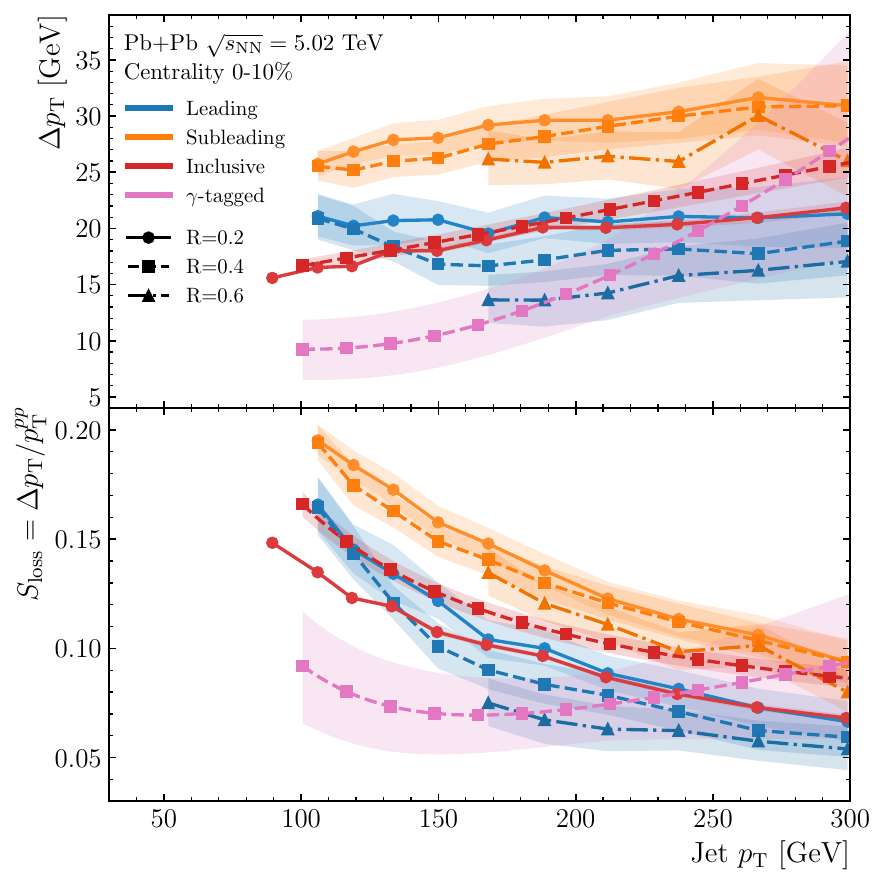}
    \caption{Leading and subleading dijets}
    \label{fig:sloss_ATLAS_leadsublead}
  \end{subfigure}
  \hfill
  \begin{subfigure}{0.49\textwidth}
    \centering
    \includegraphics[width=\textwidth]{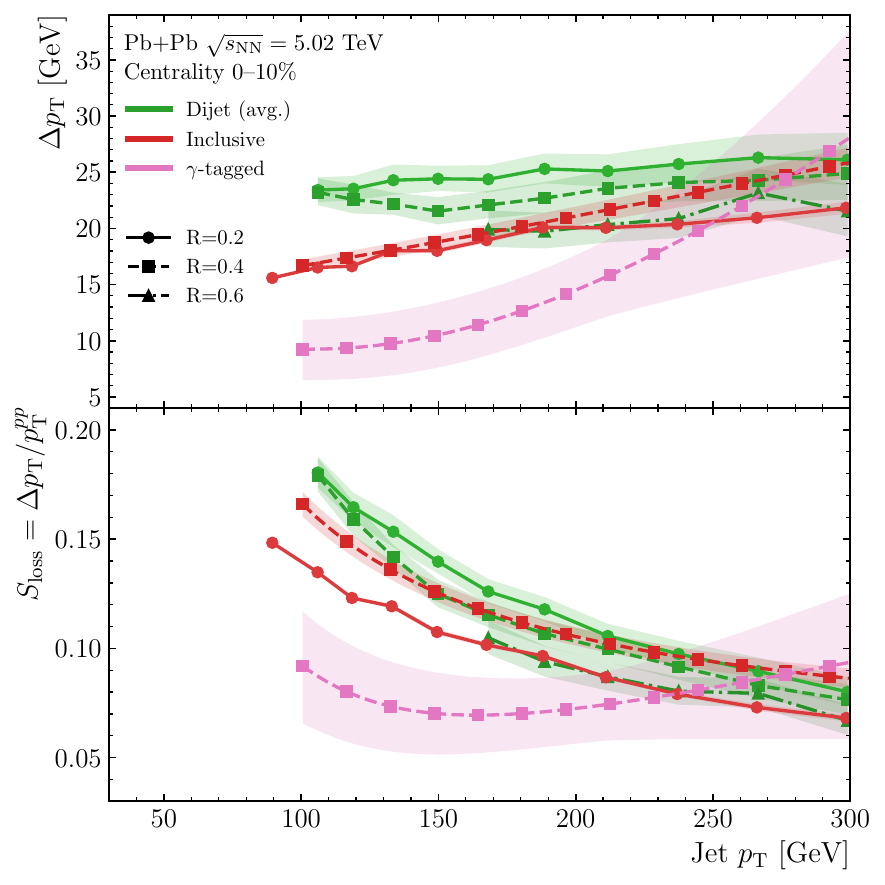}
    \caption{Dijet-averaged results}
    \label{fig:sloss_ATLAS_dijet_avg}
  \end{subfigure}
  \caption{
    Transverse momentum shift $\Delta p_{\mathrm{T}}$ (top) and fractional energy loss $S_{\text{loss}}$ (bottom) in central (0--10\%) Pb+Pb collisions at $\sqrt{s_{\mathrm{NN}}} = 5.02~\mathrm{TeV}$, extracted using matched jet spectra from ATLAS. Panel~(a) shows leading and subleading dijet results, while panel~(b) shows dijet-averaged values. Inclusive and photon-tagged jet results from ATLAS~\cite{ATLAS:2023iad} are shown for reference; their originally continuous curves are sampled here for visual consistency. Marker and line styles indicate different jet types and radii, as shown in the legend. Error bars represent combined uncertainties derived via $R_{\mathrm{AA}}$ propagation (see Sec.~\ref{sec:methods}), except for $R=0.4$ inclusive and photon-tagged jets, whose uncertainty bands are taken directly from Ref.~\cite{ATLAS:2023iad}.
    }
  \label{fig:sloss_ATLAS}
\end{figure}

Figure~\ref{fig:sloss_ATLAS} shows the transverse momentum shift $\Delta p_{\mathrm{T}}$ (top panels) and the resulting fractional energy loss $S_{\text{loss}} = \Delta p_{\mathrm{T}} / p_{\mathrm{T}}^{pp}$ (bottom panels) as a function of jet $p_{\mathrm{T}}$ for ATLAS data. Panel (a) presents results for leading and subleading jets in dijet events, while panel (b) displays dijet-averaged values. In both panels, ATLAS inclusive~\cite{ATLAS:2023iad} and photon-tagged~\cite{ATLAS:2023iad} jets (both at $R=0.4$) are shown for comparison.

To better quantify the average energy loss in dijet events, we compute the dijet-averaged quantities by taking the mean of the leading and subleading contributions at each $p_{\mathrm{T}}$ bin:
\begin{equation}
    \Delta p_{\mathrm{T}}^{\text{avg}} = \frac{\Delta p_{\mathrm{T}}^{\text{leading}} + \Delta p_{\mathrm{T}}^{\text{subleading}}}{2}, \qquad
    S_{\text{loss}}^{\text{avg}}  = \frac{S_{\text{loss}}^{\text{leading}} + S_{\text{loss}}^{\text{subleading}}}{2}.
\end{equation}
This approach provides a single representative curve for dijet energy loss, shown in Figure~\ref{fig:sloss_ATLAS_dijet_avg}, and facilitates a direct comparison with inclusive and photon-tagged jet selections.

As illustrated in Figure~\ref{fig:sloss_ATLAS_leadsublead}, subleading jets exhibit the largest momentum loss across all radii, consistent with their tendency to traverse longer path lengths in the medium and receive greater modifications. Leading jets experience intermediate energy loss, while photon-tagged jets display the smallest energy loss. This ordering is qualitatively consistent with the flavor composition of the samples: photon-tagged jets are more quark-enriched, whereas the inclusive and dijet selections are more gluon-dominated, and gluon jets are expected to undergo larger energy loss than quark jets~\cite{ATLAS:2023iad}.

A distinct difference emerges in the radius dependence across jet categories. For leading, subleading, and dijet-averaged jets, $S_{\text{loss}}$ decreases when the jet radius is increased: over most of the measured $p_{\mathrm{T}}$ range the curves follow $S_{\text{loss}}^{0.6} < S_{\text{loss}}^{0.4} < S_{\text{loss}}^{0.2}$ within uncertainties. In contrast, the inclusive jets exhibit an inverted hierarchy, with the $R=0.4$ curve lying systematically above the $R=0.2$ curve such that $S_{\text{loss}}^{0.4} > S_{\text{loss}}^{0.2}$ at all $p_{\mathrm{T}}$ values where both radii are available. For photon-tagged jets only $R=0.4$ results are available, and at that radius they show the smallest $S_{\text{loss}}$ among all jet selections up to $200~\mathrm{GeV}$.

Finally, the inclusive jet $S_{\text{loss}}$ values lie between those of the leading and subleading dijet samples for a given radius, and for $R=0.4$ the dijet-averaged $S_{\text{loss}}$ is close to the inclusive result at high $p_{\mathrm{T}}$.

\begin{figure}[hbt!]
  \centering
  \begin{subfigure}{0.49\textwidth}
    \centering
    \includegraphics[width=\textwidth]{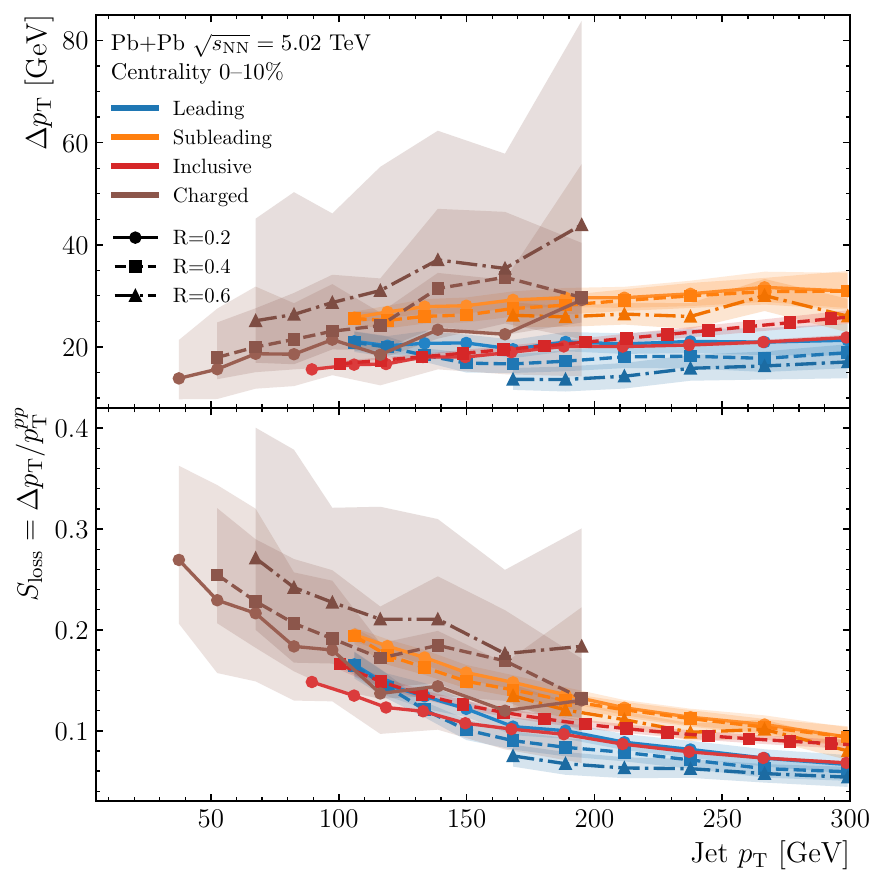}
    \caption{Leading and subleading dijets}
    \label{fig:sloss_ALICEvsATLAS_leadsublead}
  \end{subfigure}
  \hfill
  \begin{subfigure}{0.49\textwidth}
    \centering
    \includegraphics[width=\textwidth]{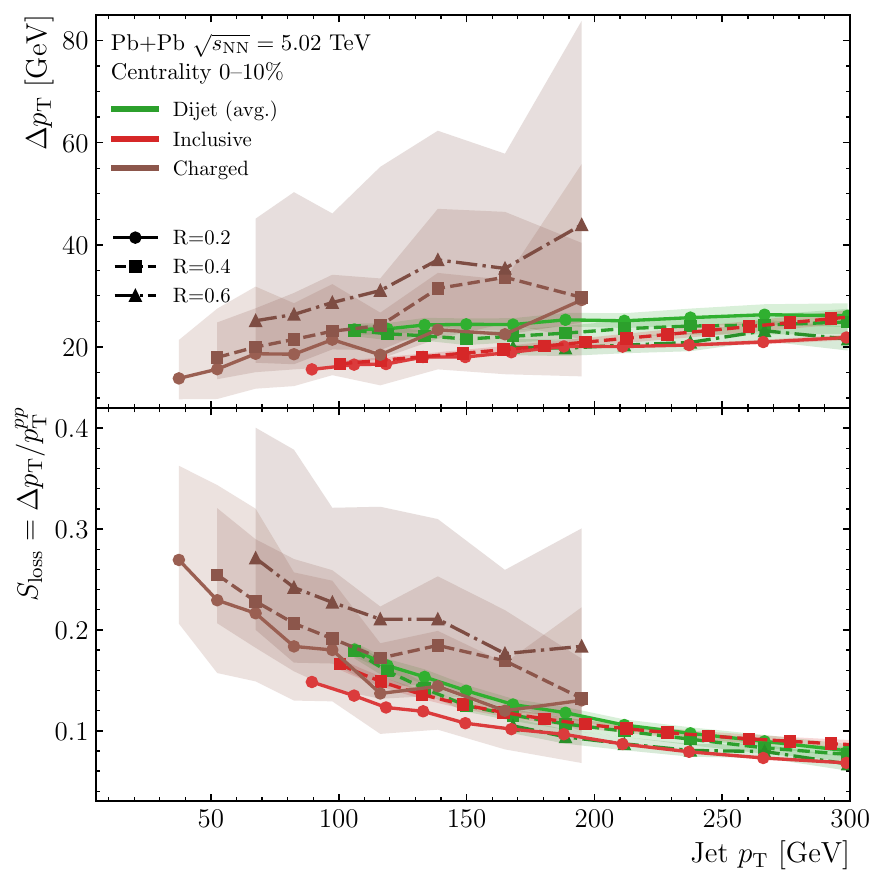}
    \caption{Dijet-averaged results}
    \label{fig:sloss_ALICEvsATLAS_dijet_avg}
  \end{subfigure}
  \caption{
    Transverse momentum shift $\Delta p_{\mathrm{T}}$ (top) and fractional energy loss $S_{\text{loss}}$ (bottom) in central (0--10\%) Pb+Pb collisions at $\sqrt{s_{\mathrm{NN}}} = 5.02~\mathrm{TeV}$. Results compare ATLAS fully reconstructed jets~\cite{ATLAS:2023hso, ATLAS:2024jtu} and ALICE charged-particle jets~\cite{ALICE:2023waz} for (a) leading and subleading dijets and (b) dijet-averaged values. Marker and line styles indicate different jet types and radii, as shown in the legend. Error bars represent combined statistical and systematic uncertainties derived from $R_{\mathrm{AA}}$ propagation (see Sec.~\ref{sec:methods}), except for ATLAS $R=0.4$ inclusive jets where uncertainty bands are taken directly from Ref.~\cite{ATLAS:2023iad}.
    }
  \label{fig:sloss_ALICEvsATLAS}
\end{figure}

To examine the consistency and differences between experiments, we compare the $S_{\text{loss}}$ values extracted from ALICE charged-particle jets to those from ATLAS fully reconstructed jets in central (0–10\%) Pb+Pb collisions. While the ATLAS results are based on calorimeter jets with full particle content, the ALICE measurements rely on charged-particle jets reconstructed with tracking information only, which can affect both the recovered jet energy and its angular structure. The extracted $\Delta p_{\mathrm{T}}$ and $S_{\text{loss}}$ for leading and subleading jets are shown in Figure~\ref{fig:sloss_ALICEvsATLAS_leadsublead}, while Figure~\ref{fig:sloss_ALICEvsATLAS_dijet_avg} presents the corresponding dijet-averaged values.

Several important trends emerge from this comparison. For $R=0.2$, the $S_{\text{loss}}$ values of ALICE charged-particle jets are comparable to those of ATLAS inclusive and leading jets, and consistently lower than the subleading dijets. For $R=0.4$, the ALICE results follow the ATLAS inclusive jet trend closely up to $p_{\mathrm{T}} \sim 120~\mathrm{GeV}$, with an upward deviation at higher momenta. The most pronounced difference appears at $R=0.6$, where the ALICE charged-particle jets exhibit systematically higher $S_{\text{loss}}$ values across the full $p_{\mathrm{T}}$ range, although the associated uncertainties are substantial, particularly in the ALICE-$p_{\mathrm{T}}$ regime. Second, the hierarchy with respect to jet radius differs between the experiments. For ALICE charged-particle jets, the extracted $S_{\text{loss}}$ values tend to increase with $R$, in contrast to the ATLAS dijet results. However, the uncertainties in the ALICE measurement are substantial, particularly at larger radius, which makes it difficult to draw a firm conclusion about the significance of this trend.

\section{Conclusions}

In this study, we performed a comparative analysis of two complementary jet-quenching observables in central Pb+Pb collisions at $\sqrt{s_{\mathrm{NN}}}=5.02~\mathrm{TeV}$: the double nuclear modification factor $R_{\mathrm{AA}}^{R/0.2}$ and the fractional energy loss $S_{\text{loss}}$.

For $R_{\mathrm{AA}}^{R/0.2}$, ATLAS leading jets in dijet events exceed unity at both $R=0.4$ and $R=0.6$, indicating reduced suppression for larger-radius jets in this selection. ATLAS subleading jets in dijet events show a smaller enhancement, while ATLAS inclusive jets exhibit only a weak radius dependence over the measured range. In contrast, ALICE charged-particle jets give $R_{\mathrm{AA}}^{R/0.2}$ values at or below unity for $R=0.4$ and $R=0.6$, indicating a different radius dependence from that observed for ATLAS calorimeter jets. Comparisons of the corresponding jet-yield ratios in $pp$ and Pb+Pb collisions show that the difference between ALICE and ATLAS is more pronounced in the Pb+Pb environment and is largest at $R=0.6$.

The extracted $S_{\text{loss}}$ values show a strong dependence on jet selection. At $R=0.4$, photon-tagged jets exhibit the smallest energy loss and subleading jets in dijet events the largest, while the inclusive and leading jets in dijet events are intermediate. For ATLAS leading, subleading, and dijet-averaged jets, $S_{\text{loss}}$ decreases with increasing jet radius, whereas the ATLAS inclusive result shows the opposite ordering between $R=0.2$ and $R=0.4$. For ALICE charged-particle jets, the extracted $S_{\text{loss}}$ values tend to increase with $R$, but the current uncertainties, particularly at larger radius, are too large to support a firm conclusion about the significance of this trend.

Taken together, these results show that the apparent radius dependence of jet quenching is strongly selection-dependent and can differ substantially between calorimeter and charged-particle jet measurements. The combined use of $R_{\mathrm{AA}}^{R/0.2}$ and $S_{\text{loss}}$ provides a broader basis for comparing jet quenching across experiments and for constraining models of in-medium jet modification.

\section{Acknowledgements}

The authors wish to thank Laura Havener for helpful discussions and insightful feedback. This research is supported by the U.S. National Science Foundation award No. 2515008. R.L. is supported by the Italian
Ministry of University and Research (MUR) through the “Rita Levi-Montalcini” Program.

\bibliography{refs}

\end{document}